\documentclass[sigconf]{acmart}
\def\BibTeX{{\rm B\kern-.05em{\sc i\kern-.025em b}\kern-.08emT\kern-.1667em\lower.7ex\hbox{E}\kern-.125emX}}

\usepackage[font=small]{caption}
\usepackage{verbatim}
\usepackage{amsmath}
\usepackage{booktabs} 
\usepackage{algorithm}
\usepackage{algcompatible}
\usepackage{algpseudocode}
\usepackage{multirow}
\usepackage{color}
\usepackage{threeparttable}
\usepackage{pifont}
\usepackage{caption}
\usepackage{subcaption}

\usepackage{tikz}
\usepackage{calc}

\copyrightyear{2020}
\acmYear{2020}
\setcopyright{acmcopyright}\acmConference[GLSVLSI '20]{Proceedings of the Great
Lakes Symposium on VLSI 2020}{September 7--9, 2020}{Virtual Event, China}
\acmBooktitle{Proceedings of the Great Lakes Symposium on VLSI 2020 (GLSVLSI '20),
September 7--9, 2020, Virtual Event, China}
\acmPrice{15.00}
\acmDOI{10.1145/3386263.3406956}
\acmISBN{978-1-4503-7944-1/20/09}

\begin{document}

%

\title{
 Effective Algorithm-Accelerator Co-design for AI Solutions on Edge Devices
}
\author{ 
Cong Hao$^{1,2}$, Yao Chen$^{3}$, Xiaofan Zhang$^{1,2}$, Yuhong Li$^{1,2}$, Jinjun Xiong$^{2,4}$,
}
\author{ 
Wen-mei Hwu$^{1,2}$, Deming Chen$^{1,2}$
}

\affiliation{ \normalsize \vspace{-8pt}
	\institution{$^1$University of Illinois at Urbana-Champaign, $^2$IBM-Illinois Center for Cognitive Computing Systems Research (C$^3$SR)}
}

\affiliation{ \normalsize \vspace{+2pt}
	\institution{ $^3$Advanced Digital Sciences Center, Singapore, $^4$IBM T. J. Watson Research Center}
}

\affiliation{ \normalsize  \vspace{+2pt}
	\textit{\{congh, xiaofan3, leeyh, w-hwu, dchen\}@illinois.edu, yao.chen@adsc-create.edu.sg, jinjun@us.ibm.com}
	 \vspace{+10pt}
}

\begin{abstract}
High quality AI solutions require joint optimization of AI algorithms, such as deep neural networks (DNNs), and their hardware accelerators.
To improve the overall solution quality as well as to boost the design productivity, 
efficient algorithm and accelerator co-design methodologies are indispensable.
In this paper, we first discuss the motivations and challenges for the Algorithm/Accelerator co-design problem,
and then provide several effective solutions.
Especially, we highlight three leading works of effective co-design methodologies:
1) the first simultaneous DNN/FPGA co-design method;
2) a bi-directional light weight DNN and accelerator co-design method;
3) a differentiable and efficient DNN and accelerator co-search method.
We demonstrate the effectiveness of the proposed co-design approaches using extensive experiments on both FPGAs and GPUs, with comparisons to existing works. 
This paper emphasizes the importance and efficacy of algorithm-accelerator co-design, and calls for more research breakthroughs in this interesting and demanding area.
\end{abstract}

\maketitle

\section{introduction}

\begin{figure}[t]
\begin{subfigure}{.45\textwidth}
  \centering
  \includegraphics[width=1.0\linewidth]{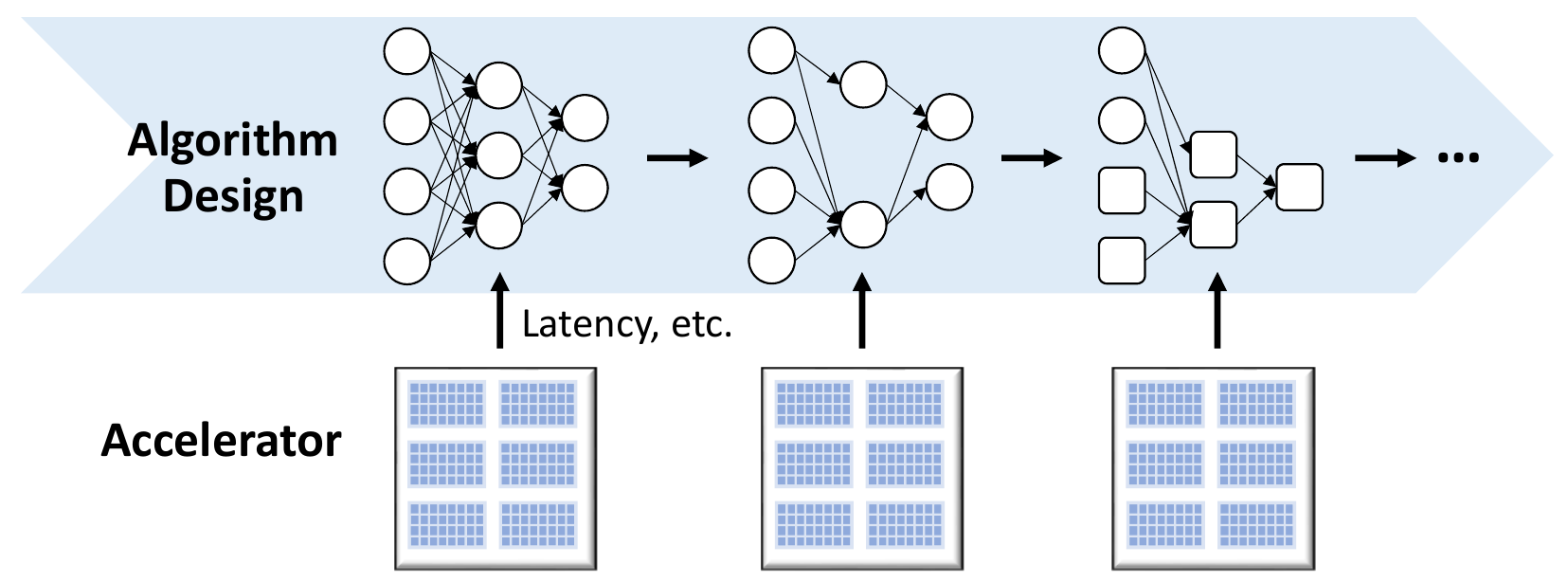}  
  \caption{Hardware-aware NAS}
  \label{fig:hw_NAS}
\end{subfigure}
\begin{subfigure}{.45\textwidth}
  \centering
  \includegraphics[width=1.0\linewidth]{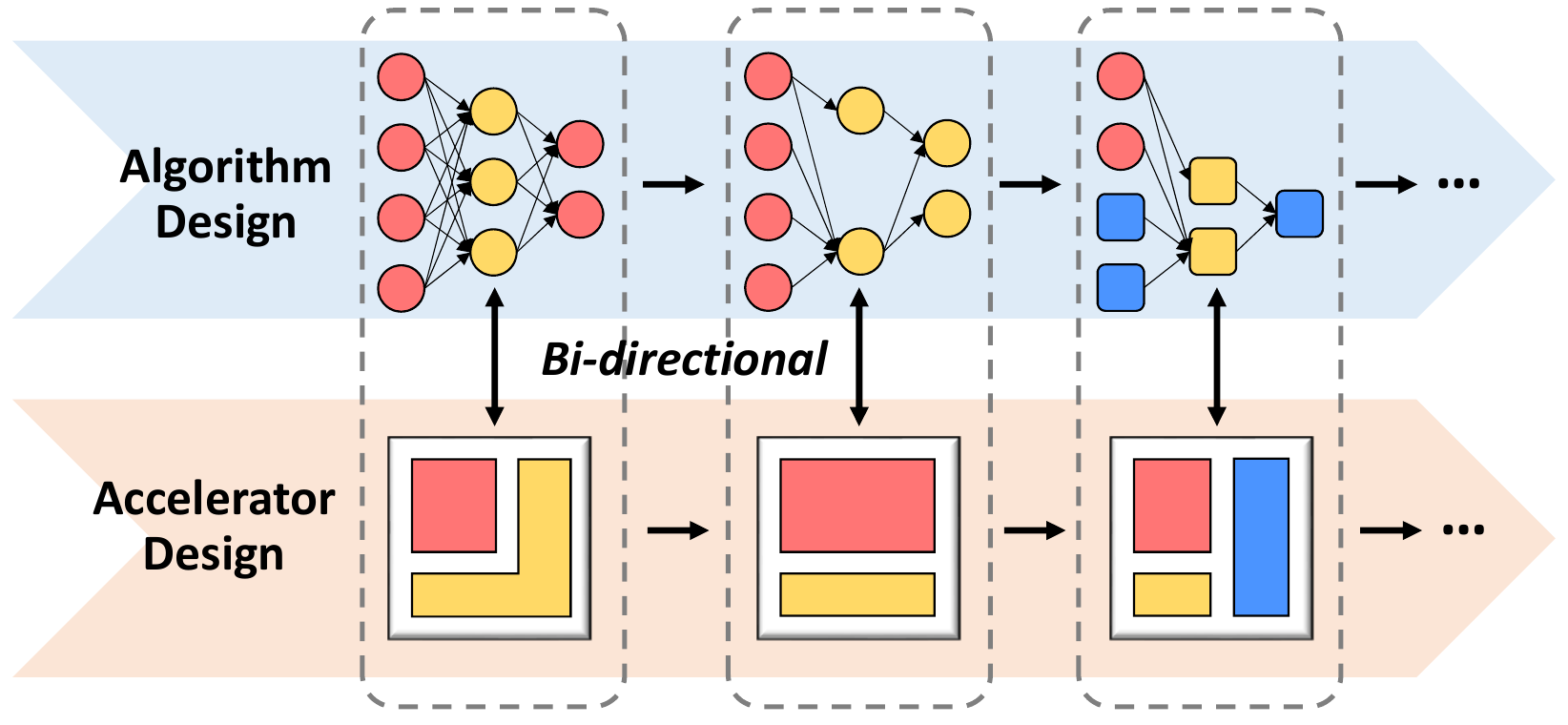}  
  \caption{Simultaneous algorithm/accelerator co-design}
  \label{fig:strict_codesign}
\end{subfigure}
\caption{(a) The general hardware-aware neural architecture search: the hardware accelerator design space is fixed. (b) The fully simultaneous algorithm/accelerator co-design: both the algorithm and accelerator design spaces are parameterized with flexibility, which should be explored and co-searched simultaneously. }
\label{fig:hw_NAS_vs_codesign}
\end{figure}

Over the past years, AI algorithms, represented by deep neural networks (DNNs), have shown significant progress in many applications including computer vision, data analytics and natural language processing (NLP), etc.
The designs in terms of the architectures and hyper parameters of DNNs have been proven to be critical for improving the quality of DNNs.
Manually designed neural network architectures have shown great performance in terms of the accuracy of the inference tasks, e.g., classification, object detection and language processing.
However, the manual designs of the algorithms typically are time consuming, error-prone and also deeply relying on human expertise, which can become the bottleneck for further improvement of DNN accuracy and compute efficiency.
Consequently, neural architecture search (NAS) is attracting more and more research interests as an alternative DNN algorithm design approach.
NAS is a powerful and promising methodology to automatically search for efficient neural networks without requiring intensive manual efforts. It can deliver DNNs with state-of-the-art performance that surpasses manual designs. 

Based on NAS, driven by the requirements of high performance hardware deployment, algorithm designers also start to explore more hardware platform friendly algorithms~\cite{cai2018proxylessnas, tan2019mnasnet, wu2019fbnet}. These approaches consider the hardware characteristics during the design of the DNN models through NAS.
Such approaches are known as \textit{hardware-aware NAS},
as shown in Figure~\ref{fig:hw_NAS}.
Hardware-aware NAS approaches try to integrate hardware performance feedback (e.g. latency) into the search process, which is usually estimated or directly measured on a \textit{fixed} hardware platform, such as GPUs or mobile phones.

Meanwhile, the optimization techniques of deploying AI algorithms on hardware platforms are also being intensively explored. These techniques include building accelerators by taking advantages of different hardware devices \cite{franklin2018nvidia,jouppi2017datacenter,isscc_2016_chen_eyeriss,zhang2017high,li2019implementing}, exploring optimization schemes to reduce the DNNs' model complexity and increase their hardware efficiency \cite{han2017ese,zhuge2018face}, and designing Electronic Design Automation (EDA) tools to enable automatic end-to-end DNN optimization and deployment \cite{wang2018design,zhang2018dnnbuilder,ye2020hybrid}.

On top of these achievements, to further improve the quality of AI solutions, AI algorithm designers and hardware developers begin to explore joint optimization opportunities, i.e., \textbf{simultaneous algorithm-accelerator co-design}.
Such an algorithm-accelerator co-design approach is different to the previous hardware-aware NAS.
As illustrated in Figure \ref{fig:strict_codesign}, 
both the algorithm and accelerator design spaces are not fixed, i.e., \textit{parameterized with flexibility}.
Thus, it requires that
the search spaces of both algorithm and accelerator to be searched in a simultaneous manner. 
As a result, the algorithm design and the accelerator design can provide instantaneous and mutual feedback and guidance to improve the solution quality on both sides.
This is especially useful when the hardware platform provides the flexibility of customizing the accelerator architecture and design.
Such a new paradigm is first proposed by Hao and Chen in \cite{hao2018deep}, and is then formally defined as \textit{NAIS -- Neural Architecture and Implementation Search} in \cite{hao2019nais}.
Following the idea in \cite{hao2018deep}, a series of co-design approaches have been proposed recently \cite{hao2018deep, hao2019fpga, hao2019nais, jiang2019accuracy,li2020edd},
among which the bi-directional approach~\cite{zhangskynet} and differentiable implementation and architecture co-search approach~\cite{li2020edd} have shown great advantages for both the neural networks and their hardware implementations.


Given the emerging requirements of algorithm-accelerator co-design, in this paper, we aim to summarize the challenges of this new methodology and present several effective co-design approaches systematically, and demonstrate a series of promising experimental results.
In Section \ref{sec:motivation} we discuss the motivations and challenges of effective algorithm-accelerator co-design.
In Section \ref{sec:related} we discuss related works.
In Section \ref{sec:solutions} we highlight three representative co-design works and their advantages.
Section \ref{sec:experiments} demonstrates promising experimental results, and Section \ref{sec:conclusion} concludes the paper and points out a few future directions.

\section{Motivations and Challenges}
\label{sec:motivation}

The AI algorithm development and accelerator deployment on hardware are two primary yet non-trivial tasks.
When targeting resource limited edge devices such as mobile and portable devices, the strict resource and power constraints and high performance requirements make both tasks more challenging.
Previous works, such as hardware-aware NAS, compact model design, model compression, and model-aware accelerator design,
attempted to resolve the conflicts between stringent hardware constraints and high model accuracy and inference speed, but these methods have drawbacks.
First, it is hard to balance between algorithm metrics (e.g. accuracy, model complexity, robustness, etc.) and hardware implementation metrics (e.g. throughput, latency, resource usage, power consumption, etc.). Thus, algorithm and hardware designers need to iterate between the algorithm and accelerator design processes, trying to help each other to achieve their own design goals. However, such an ad hoc iterative process will result in a time-consuming and error-prone procedure with no guarantee of convergence.
Second, since both tasks are not comprehensively co-designed, it can lead to sub-optimal AI algorithms (hardware-unfriendly) and accelerators (model-unfriendly). 

As a result, we have advocated that algorithms and their hardware accelerators
need to be designed simultaneously \cite{hao2018deep}\cite{hao2019nais}, which brings immense optimization opportunities.
First, co-design creates hardware-oriented or hardware-specific AI algorithms. For hardware deployment, there are many candidate hardware devices,
each of which has different characteristics in terms of 
computation capability, memory capacity and bandwidth, and reconfigurability.
The co-design methodology will search and create algorithms incorporating hardware-specific features naturally and best leveraging such hardware features for high computation efficiency.
Second, co-design meets resource and performance constraints incurred by the hardware devices. The co-design methodology will search for algorithms suitable for  available hardware resources while honoring various performance constraints, which provides predictable and guaranteed performances for hardware deployment.
Third, co-design shortens design cycles. While existing iterative design
methods require tedious efforts to find satisfying
solutions, an automated co-design flow can find an optimized AI algorithm and its efficient deployment on hardware simultaneously.

Despite such opportunities, there are many challenges as well.
First, it is hard to formulate a unified co-search space because of the very different natures of the two problems.
In addition, it is non-trivial to provide instant and proper mutual feedback during the co-design procedure, when neither the algorithm nor the accelerator architecture is known.
\textit{Thus, a proper formulation of the co-design problem is essential.}
Second, a good co-design approach requires the exploration of an extremely large number of variables in the combined algorithm and accelerator co-design space, which will be highly time-consuming and complex.
\textit{Thus, efficient search algorithms are also indispensable for rapid and effective co-search.}

\section{background and related works}
\label{sec:related}

In order to satisfy demanding AI applications, it is far from enough to optimize solely from the AI algorithm side, such as using network compression techniques to reduce the network complexity or NAS to find network models with better accuracy compared to what have been designed by human experts. Instead, we need to address challenges from both algorithm and hardware perspectives and start exploring the joint optimization opportunities between the DNNs and their hardware deployments.

The very first work that discusses algorithm/accelerator co-design is published in \cite{hao2018deep} in 2018, which introduces the opportunities and challenges of DNN and hardware accelerator co-design as well as a simultaneous co-design method to produce high-quality DNN designs and hardware implementations as a new concept. 
In 2019, the work \cite{hao2019fpga} materializes the co-design idea and creates the first simultaneous DNN/FPGA co-design framework, which helps systematically accomplishing the complicated co-design solution. This framework includes a hardware-oriented DNN model design following a bottom-up approach, and a DNN-driven FPGA accelerator design following a top-down approach. A fully automatic co-design flow is developed accordingly for simultaneous DNN search and FPGA accelerator generation.
For DNN model design, a DNN template to guide the DNN generation with predictable performance and resource utilization is proposed, which greatly reduces the co-design search space. For FPGA accelerator design, a fine-grained tile-based pipeline architecture is proposed, which supports arbitrary DNNs searched by the framework. This work demonstrates promising results on an object detection task targeting a PYNQ-Z1 embedded FPGA. DNN models are searched and mapped to the board with the state-of-the-art performance regarding accuracy, speed, and power efficiency.
Later, our work called NAIS \cite{hao2019nais} discusses the opportunities to extend the co-design framework to support both FPGA and GPU platforms. It takes autonomous driving as a case study to illustrate the importance of having comprehensive considerations of designing DNN and accelerator at the same time and demonstrates how such a co-design methodology can impact the autonomous driving industry significantly.
In order to handle scenarios where resources are more limited, SkyNet is proposed in \cite{zhangskynet} to solve challenging object detection and tracking problems using resource-constrained embedded systems. In this work,
SkyNet provides a bi-directional design strategy where light-weight DNNs are searched and designed with comprehensive understanding of the target hardware constraints, and the hardware accelerators are designed and implemented for efficient computation of the DNNs. SkyNet is demonstrated on both embedded FPGA (Ultra96) and GPU (TX2) by delivering the best object detection results regarding accuracy, throughput, and energy efficiency. In addition, SkyNet is extended to support object tracking with much less parameters but comparable accuracy when compared to the state-of-the-art trackers using ResNet-50 backbone.
On top of such achievements, the work EDD \cite{li2020edd} proposes an elegant differentiable DNN architecture and implementation co-search methodology. The co-search problem is formulated by merging DNN search variables and hardware implementation variables into one solution space, which is differentiable with respect to the merged variables, so that gradient descent algorithm can be applied to greatly reduce the search time. The formulation is also applicable to various devices with different objectives. The experiments demonstrate three representative DNNs targeting different hardware platforms with various performance requirements, each achieving the state-of-the-art accuracy with greatly improved hardware implementation performance such as latency and throughput.


There are several other DNN and accelerator co-design works, such as \cite{jiang2019accuracy, yang2020co},
demonstrating promising results and great potential to boost the quality of both machine learning algorithms and hardware implementations. In \cite{jiang2019accuracy}, a hardware-aware NAS framework called FNAS is proposed to find an optimized neural architecture under required FPGA implementation latency. With a performance model to analyze the latency of neural architectures, FNAS can prune architectures that do not satisfy the specification and generate co-design results with greatly reduced search time.
In \cite{yang2020co}, 
a framework named NASAIC is proposed to simultaneously identify multiple DNN architectures and the associated heterogeneous ASIC accelerators to meet both software accuracy and hardware performance requirements.

\section{Effective Solutions}
\label{sec:solutions}

\begin{figure}
    \centering
    \includegraphics[width=0.45\textwidth]{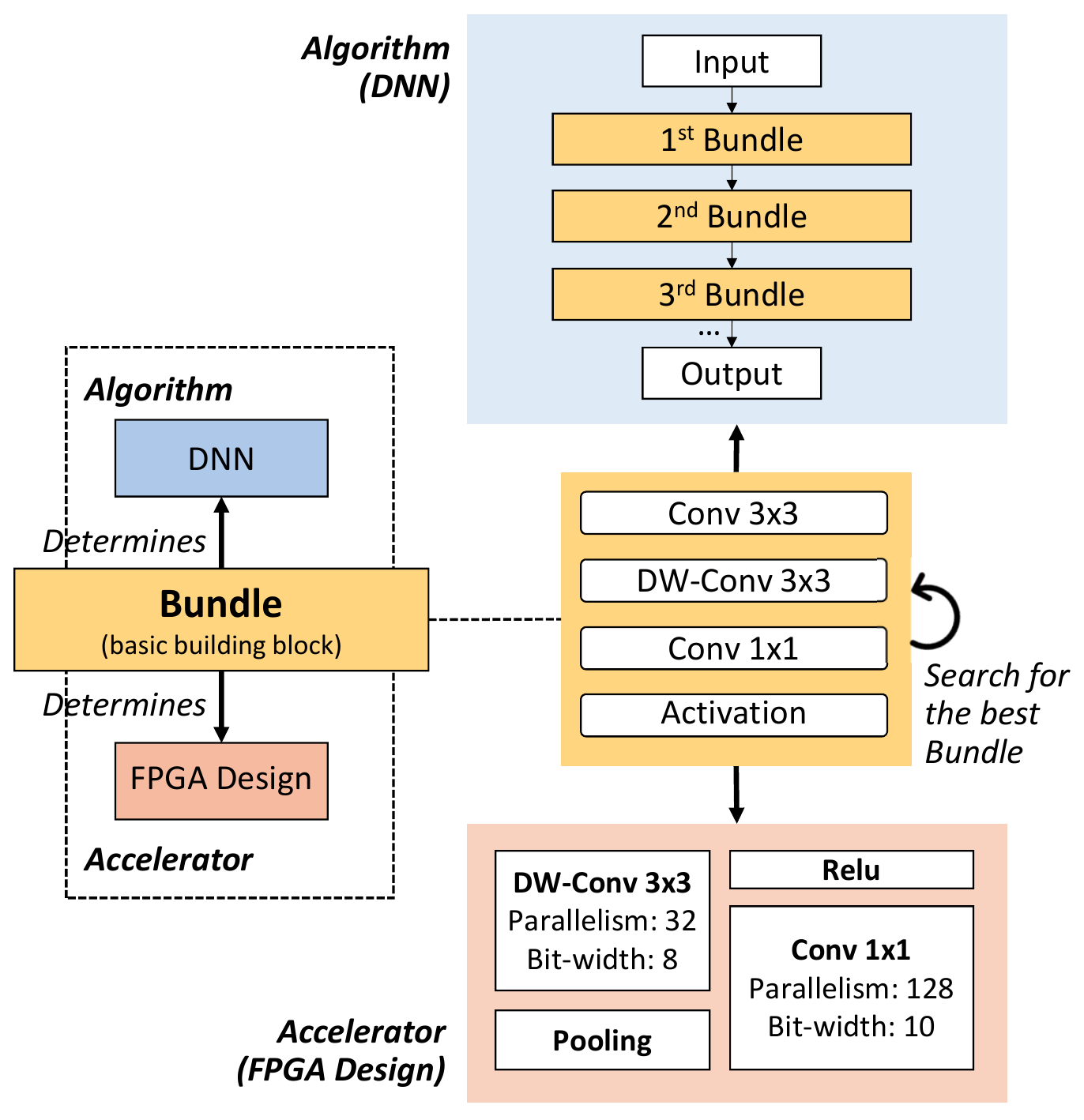}
    \caption{The basic idea of DNN/FPGA accelerator co-design \cite{hao2019fpga}.}
    \label{fig:FPGA_DNN_idea}
    \vspace{-6pt}
\end{figure}

\subsection{NAIS: Neural Architecture and Implementation Search}
After the introduction of the co-design concept in \cite{hao2018deep} and the creation of an actual co-design framework in \cite{hao2019fpga}, a formal co-design formulation called NAIS \cite{hao2019nais}, \textit{the Neural Architecture and Implementation Search}, has been defined for simultaneous algorithm-accelerator co-design in a comprehensive way, using autonomous driving as an compelling use case.
As illustrated in Figure \ref{fig:strict_codesign},
NAIS requires that \textit{both algorithm design space and accelerator design space are parameterized with flexibility}.
Following NAIS,
we proposed several effective solutions tackling the algorithm-accelerator co-design problem effectively. More details are introduced in the next several subsections.

\subsection{Simultaneous DNN/FPGA Co-design}

\begin{figure*}
    \centering
    \includegraphics[width=0.8\textwidth]{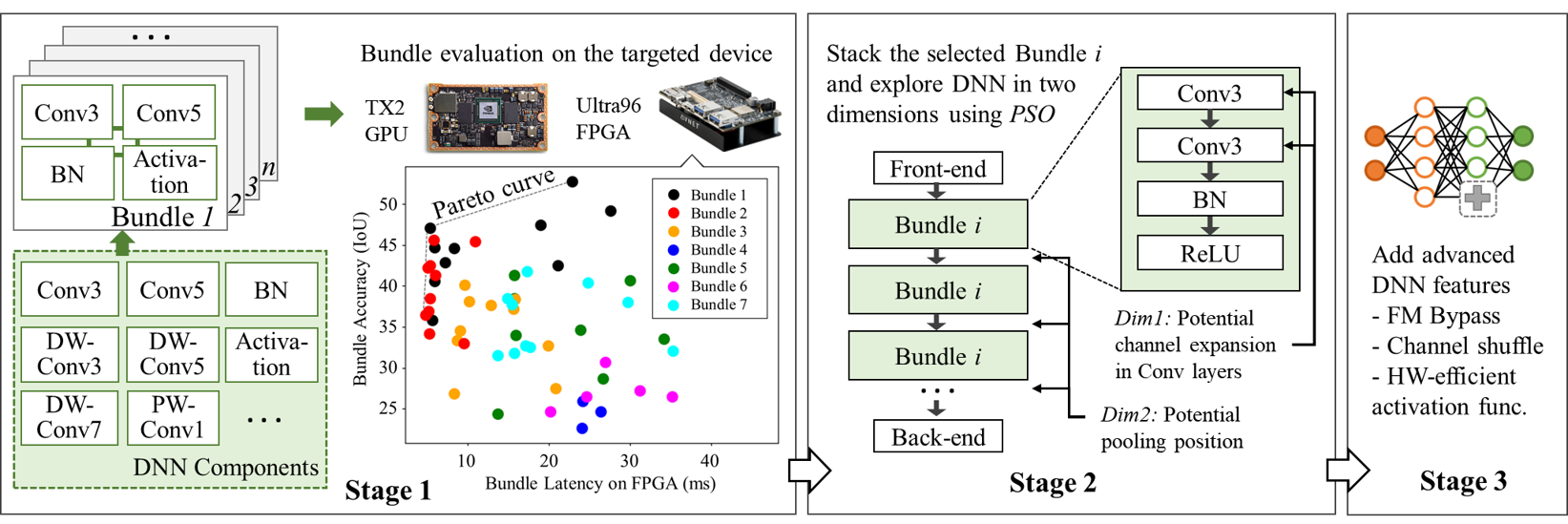}
    \vspace{-4pt}
    \caption{ SkyNet proposes a bi-directional approach for hardware-efficient DNN design \cite{zhangskynet}. }
    \vspace{-8pt}
    \label{fig:skynet_method}
\end{figure*}
As mentioned previously, the simultaneous DNN and FPGA accelerator co-design is challenging mainly because of the huge co-design space.
On the one hand, the FPGA accelerator optimization problem is by itself very complicated and requires comprehensive domain-specific knowledge, considering many different hardware design aspects, such as the overall accelerator architecture (pipelined or folded), the number of IPs and parallelism of each IP, data quantization, buffer allocation, data reuse, etc., and each has a significant impact on the final accelerator performance.
On the other hand, the neural architecture design also has a large design space, considering different architecture structures and configurations, such as the types of the layers, kernel sizes, layer dimensions, the connections between layers, etc.

To efficiently narrow down the combined design space of NAIS, 
we propose to co-design both DNN structure and its FPGA accelerator implementation using hardware-aware basic building blocks, named \textbf{Bundles} \cite{hao2019fpga}.
Figure \ref{fig:FPGA_DNN_idea} illustrates the basic idea of Bundle-based DNN and FPGA accelerator co-design.
A Bundle is the basic building block for a DNN and its corresponding FPGA accelerator, with the capability to represent both the algorithm component and the corresponding IPs to be used in the accelerator.
First, a Bundle represents a set of sequential DNN layers, 
and a DNN can be constructed by replicating a Bundle for $n$ times with configurations (the '\textbf{A}' in N\textbf{A}IS).
Meanwhile, a Bundle is composed of a set of FPGA configurable IPs, where each IP is well designed and highly optimized, and the Bundle is used to construct the FPGA implementation (the '\textbf{I}' in NA\textbf{I}S).
For DNN, each Bundle replication can be configured to have different number of channels of its layers;
for FPGA, a Bundle can be configured to have a certain number of IP instances, and each IP instance with specific parallel factors, data precision, on-chip buffers, etc.
When a Bundle is selected and configured,
both the DNN model and its accelerator can be determined. 
Thus, co-designing DNNs and FPGA accelerators equals to selecting the best Bundle and determining its configurations to construct the DNN and the accelerator.

The proposed DNN/FPGA accelerator co-design has three major steps,
in order to select the best Bundle and determine the configurations.
The inputs include a machine learning task such as image classification or object detection, resource constraints of a specific FPGA device, and performance target such as frame rate.
The outputs include both DNN models and corresponding FPGA accelerator with achieved performance.

\textbf{Step 1: FPGA-oriented Bundle generation}.
First, a pool of FPGA-oriented IPs considering specific FPGA characteristics such as DSP and BRAM structures is designed.
The IPs may have same functionality but different designs.
For example, to best utilize the DSP resource, a Xilinx FPGA may best support 8-bit $\times$ 10-bit multiplication IPs, while an Intel FPGA may best support 9-bit $\times$ 9-bit multiplication IPs.
Based on the IPs, we build FPGA-oriented Bundles, where the data tiling, pipelining and data movement between these IPs are considered.
Given DNN building blocks and a hardware IP pool, analytical models are built to capture the hardware latency and resource utilization of the building blocks and the DNNs built from the blocks. This is to provide performance estimation in the early stage of DNN exploration.
    
\textbf{Step 2: Bundle selection}.
Second, Bundles are evaluated to reduce the co-design space by only selecting the most promising ones for future exploration.
Each Bundle will be evaluated regarding its resource utilization and potential contribution to DNN accuracy.
We build a Bundle-wise DNN template with fixed front-end and back-end structures, and insert one Bundle (with replications) in the middle each time. Such Bundle-wise DNNs will be quickly trained using a small number of epochs to evaluate the accuracy.
The Bundles on the resource-accuracy Pareto curve will be selected.
 
\textbf{Step 3: Hardware-aware DNN search and update}. 
Third, DNNs are searched and fine-tuned using the selected promising Bundles within the performance requirements and resource constraints. The inputs include the initial DNNs and performance objectives (e.g., latency, resource constraints). The stochastic coordinate descent (SCD) is used to update DNN construction related variables, including the number of Bundle replications, down-sampling configuration between Bundles, and channel number in each Bundle. During the iterations of SCD, only DNNs within the resource constraints and performance requirements are kept for downstream training.
In such a way, the final generated DNNs are more structured, resulting in more efficient hardware implementations.

\subsection{Bi-directional DNN/Accelerator Co-design}

\begin{figure*}
    \centering
    \vspace{-8pt}
    \includegraphics[width=0.75\textwidth]{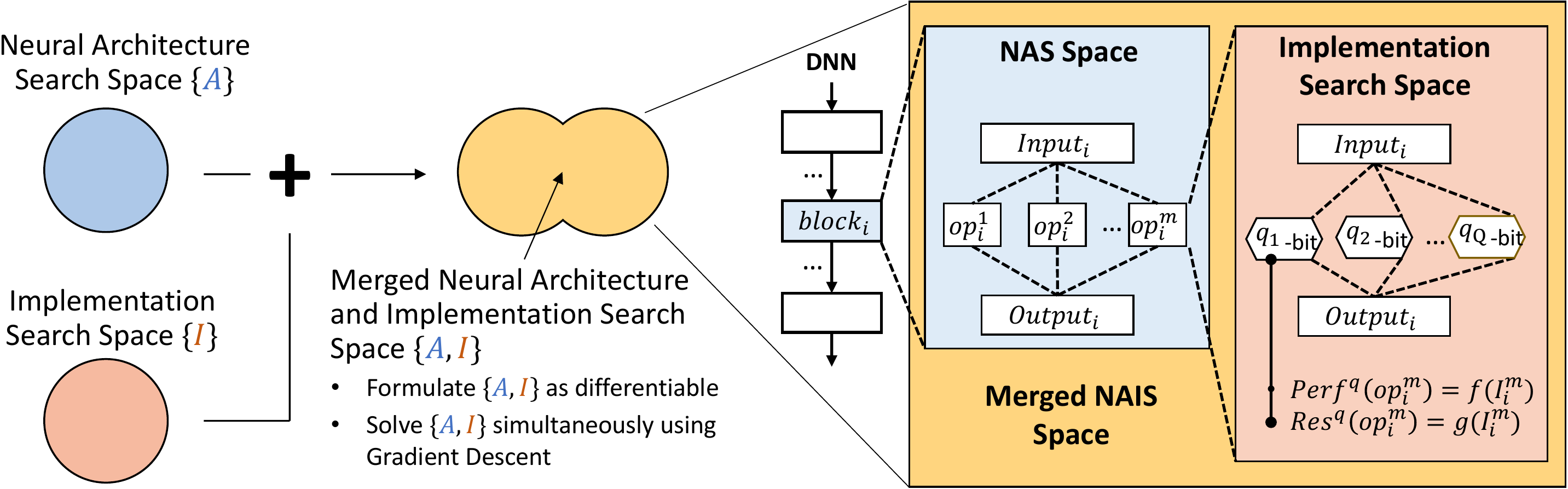}
    \caption{The basic idea of EDD \cite{li2020edd}: merging two design spaces into a single space which is differentiable with respect to design variables. \vspace{-8pt}}
    \label{fig:EDD_idea}
\end{figure*}
Based on the FPGA/DNN co-design approach \cite{hao2019fpga} which targets FPGA accelerators only, SkyNet \cite{zhangskynet} is further proposed as a bi-directional approach, to overcome challenges while designing hardware efficient DNNs for embedded FPGAs and GPUs. The proposed bi-directional design method is summarized in Figure \ref{fig:skynet_method}.

The proposed method starts by building the Bundle to capture the hardware resource constraints and the realistic latency results for both embedded FPGAs and GPUs, which is similar to \cite{hao2019fpga}.
An improvement compared to \cite{hao2019fpga} is that,
after Bundle analysis, SkyNet introduces a novel DNN search stage to select top candidates which can meet the requirements from both hardware and software perspectives. Since multiple metrics need to be considered, the candidates search can be summarized as solving a multi-objective optimization problem. A particle swarm optimization (PSO) evolutionary algorithm is proposed to discover proper DNNs. In this algorithm, each individual DNN is regarded as a particle, and all active DNNs during the search contribute to the swarm, where DNNs composed by the same type of Bundle are considered as in the same particle group. 
A fitness value is defined to evaluate each particle covering both DNN accuracy and hardware latency information when deploying on targeted hardware devices. Then, the global optimal and the group optimal designs are kept to provide evolutionary directions for each particle. We also enable two hyper-parameters, which act as the variable factors for each particle, including the number of channels of each Bundle replication and the pooling position between Bundles. Both of these variable factors affect accuracy and hardware performance. 
During search, the updated velocity of a particle (which represents a DNN candidate) is calculated during every iteration based on the current velocity.
Particles can move to a better position following the predefined policy.
Eventually, the best DNN in the global best position will be selected.
Finally, more advanced DNN design features are added to further improve the DNN accuracy and hardware efficiency. For the object detection and tracking tasks, a layer-bypass structure from low-level to high-level layers is added, involving feature map reordering to strengthen the capability of small object detection. Also, the standard ReLU layer is replaced by the ReLU6 for better hardware efficiency.

\subsection{Differentiable DNN/Accelerator Co-design}


On top of the co-design approach \cite{hao2019fpga} targeting FPGA only, and the SkyNet bi-directional approach \cite{zhangskynet} targeting GPU and FPGA, we further extend NAIS to a unified co-design space, aiming to systematically formulate the co-design problem with a more efficient solution.
Our proposed solution is called EDD \cite{li2020edd}, an elegant and \textbf{E}fficient \textbf{D}ifferentiable \textbf{D}eep neural architecture and implementation co-search formulation and methodology.

Figure \ref{fig:EDD_idea} illustrates the EDD idea.
There are two design spaces, a DNN architecture search space $A$, depicted as blue,
and an implementation search space $I$, depicted as brown.
The two spaces are merged into a single co-search space $\{A, I\}$, depicted as yellow, with a joint objective function $\mathcal{L} = \mathcal{F}(A, I)$.
Moreover, EDD proposes a \textit{differentiable approach} for solving the merged $\{A, I\}$ based on two main considerations.
First, by formulating $\mathcal{L}$ as differentiable with respect to both $A$ and $I$, $A$ and $I$ can be seamlessly optimized simultaneously using optimization algorithms over a continuous space, such as gradient descent.
Second, among all NAS approaches searching for $A$, differentiable NAS, such as DARTS \cite{liu2018darts} and FBNet\cite{wu2019fbnet}, has been proven to be GPU-hour efficient with appealing model accuracy.

Consequently, the key challenge is how to formulate $A$ and $I$ as \textit{continuous} variables.
First, for DNN search, the goal is to quickly find a DNN architecture while minimizing accuracy loss, denoted as $Acc_{loss}$.
Second, for implementation search, a \textit{performance loss} is defined and is denoted as $Perf_{loss}$, which can be specified by users, such as end-to-end inference latency, throughput, energy, DNN model complexity, etc. The resource utilization is denoted as $RES$, and the resource upper-bound of the target hardware as $RES_{ub}$.
Therefore, the objective of EDD is to minimize accuracy loss $Acc_{loss}$ and performance loss $Perf_{loss}$ simultaneously by effectively searching $\{A, I\}$:
\begin{equation}
    min:\mathcal{L} = Acc_{loss}(A,I) \cdot Perf_{loss}(I) + \beta \cdot C^{RES(I) - RES_{ub}}
    \label{eq:co_search_obj_func}
\end{equation}

In Eq.~\ref{eq:co_search_obj_func}, $Acc_{loss}$ is a function of $A$ and $I$;
$Perf_{loss}$ and $RES$ are functions of $I$.
Resource upper-bound $RES_{ub}$ is expressed in an exponent term to introduce large penalty when being violated.
Worth noting, in the existing hardware-aware NAS approaches, only $A$ is searched while $I$ is \textit{fixed} during NAS;
while in EDD co-search formulation, $I$ is also \textit{variable}.
$Acc_{loss}$ is differentiable with respect to $A$ and $I$, and
$Perf_{loss}$ and $RES(I)$ are differentiable with respect to $I$.
By descending $\mathcal{L}$ with respect to the variables in $\{A, I\}$ on validation set as $\bigtriangledown_{\{A, I\}}\mathcal{L}_{val}$, $\{A,I\}$ will be searched simultaneously.


\textbf{Differentiable Neural Architecture Search Space}.
The differentiable NAS space is inspired by \cite{liu2018darts}.
First, the DNN is composed of $N$ basic building blocks, $block_i$, where $1\leq i \leq N$.
In order to design hardware-friendly DNNs and to reduce search time, EDD adopts the 
single-path DNN structure without branches \cite{stamoulis2019single}.
Comparing to the complicated layer connections and branches as in \cite{liu2018darts} \cite{tan2019mnasnet}, the single-path structure significantly reduces off-chip data transfer without sacrificing accuracy \cite{stamoulis2019single}.
Inside the $i$-$th$ block, there are $M$ candidate operations, denoted as $op_i^m$ ($1 \leq m \leq M$).
The operations are the most commonly used DNN blocks in NAS approaches, called MBConv~\cite{tan2019mnasnet}. It is composed of sequential layers of $conv$-$1\times1$, $dwconv$-$k\times k$ (depth-wise convolution with kernel size $k$) and $conv$-$1\times1$.
Between $conv$-$1\times1$ and $dwconv$-$k\times k$, the number of channels expands/shrinks by a ratio of $ch_i^m$.
The output of a block is calculated based on the outputs of its $M$ candidate operations. For example in \cite{liu2018darts}, the output is the weighted sum of the $M$ operations, where the weights are determined by a Softmax function.
Instead of Softmax, EDD adopts the Gumbel-Softmax function in \cite{wu2019fbnet} in order to sample
only one operation out of $M$ during feedforward propagation,
since Gumbel-Softmax function can convert the discrete non-differentiable sampling to continuous differentiable sampling.
This greatly reduces the memory requirement and speeds up the feedforward propagation.
The sampling parameters $\theta_{i,m}$ organize a two-dimension $N\times M$ array, denoted as $\Theta \in A$, which is the primary DNN search variable.

\textbf{Differentiable Implementation Search Space}.
The implementation search space also has to be formulated as continuous and differentiable.
During NAS, each candidate operation $op_i^m$ has its own implementation variables, forming an implementation search space $I_i^m$.
Data precision (quantization) $q$ is one of the important implementation variables, since it has a large impact on DNN accuracy, implementation performance and hardware resource.
Rather than a train-and-quantize manner, the quantization shall be searched together with DNN structure to provide implementation performance feedback.
Besides quantization, other implementation variables may be device oriented. For example, FPGA implementation design space includes parallelism, loop tiling factors, etc.
Based on Eq. \ref{eq:co_search_obj_func}, the implementation performance related terms, $Perf_{loss}$ and $RES$, shall be formulated using implementation variables.

First, data quantization is considered. 
Inspired by \cite{wu2019fbnet},
to enable differentiable quantization formulation, $Q$ quantization paths are built for each operation $op_i^m$, indicating each operation has $Q$ quantization choices, from $q_1$-$bit$ to $q_Q$-$bit$.
Similar to differentiable NAS formulation, each quantization scheme is also sampled by the Gumbel-Softmax function with a sampling parameter $\phi_{i, m, q}$, generating a possibility for $op_i^m$ to be quantized to $q$-$bit$.
The $\phi_{i, m, q}$ organizes a three-dimension array of size $N\times M\times Q$, denoted as $\Phi$.
In this formulation, EDD has the flexibility to choose different quantizations for different layers of a DNN; such a mixed precision computation can be well supported by reconfigurable hardware and dedicated accelerators \cite{sharma2018bit, sharify2018loom}. 

Second, for each operator $op_i^m$ under $q$-bit quantization, we consider other implementation variables, denoted as $I_i^m$, which are used to formulate the accelerator performance and resource utilization,
denoted as $Perf^q(op_i^m)=f(I_i^m)$ and $Res^q(op_i^m)=g(I_i^m)$, respectively.
Consequently, 
$Perf^q(op_i^m)$ and $Res^q(op_i^m)$ must be differentiable with respect to $I_i^m$.
Since the formulation of $Perf^q(op_i^m)$ and $Res^q(op_i^m)$ are highly platform dependent,
we use FPGA as an example for demonstration,
and introduce \textit{parallel factors} of the processing elements of an FPGA accelerator, denoted as $pf_i^m$.
Parallel factors describe the \textit{parallelism}, indicating how many multiplications can be done concurrently.
In FPGA design, since the parallelism usually increases exponentially such as 64, 128, 256, etc., the exponential form of $2^{pf}$ is adopted to describe parallelism.
Under such assumption, the $Perf^q(op_i^m)$ and $Res^q(op_i^m)$ can be expressed as differentiable with respect to $pf_i^m$. 

Finally, using the quantization variables and other implementation variables such as $pf$,
the final performance loss of the DNN ($Perf_{loss}$) and the resource utilization ($RES(I)$) in Eq.~\ref{eq:co_search_obj_func} can be formulated stage by stage. More details could be found in \cite{li2020edd}.

\vspace{-8pt}
\section{Experimental Results}
\label{sec:experiments}

In this section we present the experimental results of our co-design methodologies.

We first demonstrate the DNN and accelerator co-design approaches including the work \cite{hao2019fpga} and SkyNet \cite{zhangskynet} on an object detection challenge from the system design competition affiliated with IEEE/ACM Design Automation Conference (DAC-SDC) \cite{xu2019dac}.
The challenge targets the single object detection for drones with comprehensive evaluation system considering design accuracy, throughput, and energy consumption. The goal of this competition is to provide unified edge-platforms to develop and compare state-of-the-art object detection system designs.
The hardware platforms include Pynq-Z1 FPGA (2018), Nvidia TX2 GPU (2018, 2019) and Ultra96 FPGA (2019).

Table \ref{tab:rst_comp_fpga} summarizes the results where the co-design approaches in \cite{hao2019fpga} and \cite{zhangskynet} are applied, together with the results of other participants in this contest.
We show the results of different platforms together, aiming to provide a better understanding of the accelerator's different performance metrics, such as IoU (intersection over union), FPS (frame per second), power and energy efficiency (J/pic).
It first shows that SkyNet \cite{zhangskynet} outperforms all other designs on GPU by delivering the highest IoU, 73.1, and the highest speed, 67.33 FPS.
It also achieves an IoU of 71.6 on Ultra96 FPGA, where the accuracy drop is mainly because of fixed point data quantization.
Second, it shows that the DNN/FPGA co-design work \cite{hao2019fpga} delivers the best energy efficiency on a Pynq-Z1 FPGA, with energy efficiency from 0.08 J/pic to 0.14 J/pic, while still maintains high accuracy.
Comparing to the SystemETHZ solutions \cite{systemsETHZ, systemsETHZ19} which also achieve remarkable energy efficiency of 0.09 J/pic and 0.12 J/pic,
the accuracy of \cite{hao2019fpga} is more than 10\% higher.
Finally, we can observe that most FPGA designs excel GPU designs in terms of energy efficiency.

\begin{table*}[t]
    \centering
    \vspace{-8pt}
    \caption{Evaluations of our proposed co-design approaches \cite{hao2019fpga, zhangskynet} on the object detection task in the Design Automation Conference System Design Contest (DAC-SDC) \cite{xu2019dac}.}
    \newcommand{\tabincell}[2]{\begin{tabular}{@{}#1@{}}#2\end{tabular}}

    \begin{tabular}{c|c|c|c|c|c|c}
    \hline
    
    & Year & Platform & Accuracy (IoU \%)   & Speed (FPS) & Power (W) & Energy Efficiency (J/pic) \\ \hline
    \multirow{2}{*}{SkyNet \cite{zhangskynet} }
    & 2019 & Nvidia TX2 GPU & \textbf{73.1} & \textbf{67.33} & {13.50} & {0.20} \\
    & 2019 & Ultra96 FPGA   & 71.6 & 25.05 & 7.26  & 0.29  \\

    \hline
    \multirow{3}{*}{{\tabincell{c}{FPGA/DNN Co-design\\ (3 DNNs) \cite{hao2019fpga} }} } 
    & 2018 & Pynq-Z1 FPGA & 68.6 & 17.4 & 2.5 & \textbf{0.14} \\
    & 2018 & Pynq-Z1 FPGA & 61.2 & 22.7 & \textbf{2.4} & \textbf{0.11} \\
    & 2018 & Pynq-Z1 FPGA & 59.3 & 29.7 & \textbf{2.4} & \textbf{0.08} \\
     
    \hline
     
     Thinker \cite{thinker} 
     & 2019 & Nvidia TX2 GPU & 71.3 & 28.79 & 8.55 & 0.30  \\
     
     DeepZS \cite{DeepZS} 
     & 2019 & Nvidia TX2 GPU & 72.3 & 26.37 & 15.12 & 0.57 \\
     
     ICT-CAS \cite{ICT-CAS} 
     & 2018 & Nvidia TX2 GPU & 69.8 & 24.55 & 12.58 & 0.51 \\
     
     SDU-Legend \cite{SDU-Legend} 
     & 2018 & Nvidia TX2 GPU & 68.5 & 23.64 & 10.31 & 0.44 \\
     
     XJTU\_Tripler \cite{XJTU} 
     & 2019 & Ultra96 FPGA & 61.5 & 50.91 & 9.25 & 0.18 \\
     
     SystemsETHZ \cite{systemsETHZ19} 
     & 2019 & Ultra96 FPGA & 55.3 & 55.13 & 6.69 & 0.12 \\
     
     SystemsETHZ \cite{systemsETHZ} 
     & 2018 & Pynq-Z1 FPGA & 49.2 & 25.97 & 2.45 & 0.09 \\
     
     TGIIF \cite{TGIIF} 
     & 2018 & Pynq-Z1 FPGA & 62.4 & 11.96 & 4.20 & 0.35 \\
     
     iSmart2 \cite{iSmart2} 
     & 2018 & Pynq-Z1 FPGA & 57.3 & 7.35 & 2.59 & 0.35 \\
     
     \hline
    
    \end{tabular}
    \label{tab:rst_comp_fpga}
\end{table*}

\begin{table}[t]
\small
\centering
\caption{Performance of SiamRPN++ trackers on GOT-10k with different backbones evaluated on single Nvidia 1080Ti \cite{zhangskynet}.} 
\label{tab:siamrpngot10k}
\begin{tabular}{c|c|c|c|c}
\hline
 Backbone & $AO$ & $SR_{0.50}$ & $SR_{0.75}$ & $FPS$ \\ \hline
AlexNet & 0.354 & 0.385 & 0.101 & 52.36\\
ResNet-50 & 0.365 & 0.411 & 0.115 & 25.90\\
\textbf{SkyNet \cite{zhangskynet}} & 0.364 & 0.391 & 0.116 & 41.22\\
\hline
\end{tabular}
\end{table}

\begin{table}[t]
\small
\centering
\caption{Performance of SiamMask trackers on GOT-10k with different backbones evaluated on single Nvidia 1080Ti \cite{zhangskynet}.} \label{tab:siammaskgot10k}
\begin{tabular}{c|c|c|c|c}
\hline
Backbone & $AO$ & $SR_{0.50}$ & $SR_{0.75}$ & $FPS$ \\ \hline
ResNet-50 & 0.380 & 0.439 & 0.153 & 17.44\\
\textbf{SkyNet \cite{zhangskynet}}  & 0.390 & 0.442 & 0.158 & 30.15\\
\hline
\end{tabular}
\end{table}

\begin{table}[t]
    \centering
    \renewcommand{\arraystretch}{0.90}
    \small
    \caption{Comparisons with existing NAS solutions \cite{li2020edd}.}
    \label{tab:nas_comp}
    \vspace{-8pt}
    \setlength{\tabcolsep}{2pt}
    \begin{tabular}{c|c|c|c|c}
    \hline
    & \multicolumn{2}{c|}{Test Error (\%)} & GPU Latency & FPGA Latency\\ 
    \hline
    & Top-1 & Top-5 & Titan RTX & ZCU102 \cite{CHaiDNN}\\
    \hline
    \multicolumn{5}{l}{\textbf{Baseline Models}}\\
    \hline
    GoogleNet  & 30.22 & 10.47 & 27.75 ms & 13.25 ms \\
    MobileNet-V2 \cite{sandler2018mobilenetv2}  & 28.1 & 9.7 & 17.87 ms & 10.85 ms\\
    ShuffleNet-V2 \cite{ma2018shufflenet} & 30.6 & 11.7 & 21.91 ms & NA \\
    ResNet18  & 30.2 & 10.9 & 9.71 ms & 10.15ms \\
    \hline
    \multicolumn{5}{l}{\textbf{Hardware-aware NAS Models}}\\
    \hline
    MNasNet-A1 \cite{tan2019mnasnet}  & 24.8 & 7.5 & 17.94 ms & 8.78 ms\\
    FBNet-C \cite{wu2019fbnet} & 24.9 & 7.6 & 22.54 ms & 12.21 ms\\
    Proxyless-cpu \cite{cai2018proxylessnas} & 24.7 & 7.6 & 21.34 ms & 10.81 ms \\
    Proxyless-Mobile \cite{cai2018proxylessnas} & 25.4 & 7.8 & 21.23 ms & 10.78 ms \\
    Proxyless-gpu \cite{cai2018proxylessnas} & 24.9 & 7.5 & 15.72 ms & 10.79 ms \\
    \hline
    \textbf{EDD-Net-1 \cite{li2020edd}} & 25.3 & 7.7 & \textbf{11.17 ms} & 11.15 ms \\
    \textbf{EDD-Net-2 \cite{li2020edd}} & 25.4 & 7.9 & 13.00 ms & \textbf{7.96 ms} \\

    \hline
    \end{tabular}
    \\
\end{table}

\begin{table}[t]

    \centering
    \renewcommand{\arraystretch}{0.9}
    \small
    \caption{EDD-Net-1 \cite{li2020edd} accuracy and latency on 1080 Ti .}
    \label{tab:edd-net-1}
    \vspace{-8pt}
    \setlength{\tabcolsep}{2pt}
    \begin{tabular}{c|c|c |c}
    \hline
    & 32-bit Floating & 16-bit Floating & 8-bit Integer \\\hline
    Test Error & 25.5\% & 25.3\% & 26.4\% \\ \hline
    Latency & 2.83 ms &  2.29 ms & 1.74 ms \\

    \hline
    \end{tabular}
\end{table}

\begin{table}[t]

    \centering
    \renewcommand{\arraystretch}{0.9}
    \small
    \caption{Comparison between EDD-Net-3 \cite{li2020edd} and DNNBuilder\cite{zhang2018dnnbuilder}}
    \label{tab:edd-net-3}
    \vspace{-8pt}
    \setlength{\tabcolsep}{2pt}
    \begin{tabular}{c|c|c |c}
    \hline
     & Top-1 Error (\%)  & Top-5 Error (\%) & Throughput (ZC706) \\\hline
    VGG16 & 29.5 & 10.0 & 27.7 fps \\\hline
    EDD-Net-3 & 25.6 & 7.7 & 40.2 fps\\

    \hline
    \end{tabular}
    \vspace{-4pt}
\end{table}

Since SkyNet can deliver real-time object detection on embedded systems, we setup experiments on the GOT-10k benchmark \cite{huang2018got} to demonstrate its potential on object tracking. 
GOT-10k is a large high-diversity database for generic object tracking with rich motion trajectory and wide coverage of object classes. 
Models are evaluated with two metrics in GOT-10k as average overlap (AO) and success rate (SR). AO is defined as the mean of IoU between prediction and ground truth bounding boxes, while SR is defined as the proportion of predictions where the IoU is beyond some threshold. 
During evaluation, Got-10K only provides the ground truth bounding box in the first frame and expect trackers to keep tracking on the same object for subsequent frames by predicting bounding boxes.
The predictions will then be evaluated by the Got-10K server. 
We integrate the SkyNet backbone with two of the state-of-the-art trackers (SiamRPN++ \cite{li2018siamrpn++} and SiamMask \cite{wang2019fast}) and evaluate its capability of real-time tracking.
SiamRPN++ \cite{li2018siamrpn++} is the first Siamese tracker that has been proven to profit from using DNN backbones with different capacities as long as they are properly trained. To evaluate the performance of different backbones, we train three SiamRPN++ trackers with AlexNet, ResNet-50, and SkyNet backbones on GOT-10k, and the results are shown in Table \ref{tab:siamrpngot10k}. It shows that SkyNet achieves nearly the same quality (AO and SR) as the ResNet-50 backbone but much better speed (1.59$\times$ faster).
SiamMask \cite{wang2019fast} is another Siamese tracker which outperforms SiamRPN++ by incorporating image segmentation for object tracking tasks. Using SiamMask tracker, the SkyNet backbone outperforms ResNet-50 in all metrics with better tracking quality and 1.73$\times$ speedup, as shown in Table  \ref{tab:siammaskgot10k}.

In addition, the effectiveness of the EDD method is tested on both GPU and FPGA platforms.
Three DNNs are searched:
EDD-Net-1 targets GPU platform for low latency implementation;
EDD-Net-2 targets recursive FPGA accelerators for low latency;
EDD-Net-3 targets pipelined FPGA accelerators for high throughput.
Each model is produced through EDD within a 12-hour search on a P100 GPU.
EDD-Net-1 is compared with the state-of-the-art hardware-aware NAS approaches as shown in Table \ref{tab:nas_comp}, where the GPU latency is tested on Titan RTX.
The EDD-Net-1 achieves the shortest latency of 11.17ms comparing with the state-of-the-art DNN models and other mobile-oriented NAS results with similar accuracy on ImageNet. EDD-Net-1 is 1.4$\times$ faster than Proxyless-GPU \cite{cai2018proxylessnas},  2.0$\times$ faster than FBNet-C \cite{wu2019fbnet} and 1.6$\times$ faster than MNasNet \cite{tan2019mnasnet}. Table \ref{tab:edd-net-1} shows the accuracy and latency results of EDD-Net-1 on Nvidia 1080 Ti GPU after re-training and fine-tuning using TensorRT.
EDD-Net-2 is evaluated using the well-recognized CHaiDNN framework~\cite{CHaiDNN}, which is a recursive FPGA accelerator capable of executing provided DNN tasks.
The FPGA latency is collected by running the DNN models with CHaiDNN accelerators on a ZCU102 FPGA as shown in Table \ref{tab:nas_comp}, where ShuffleNet \cite{ma2018shufflenet} is currently not supported by CHaiDNN.
EDD-Net-2 delivers the shortest latency on FPGA among all the DNNs, 7.96 ms, which is 1.37$\times$ faster than ProxylessNet \cite{cai2018proxylessnas}, 1.53$\times$ faster than FBNet \cite{wu2019fbnet} and 1.1$\times$ faster than MNasNet \cite{tan2019mnasnet}.
EDD-Net-3, targeting a pipelined FPGA accelerator, is compared to the state-of-the-art accelerator design proposed by DNNBuilder \cite{zhang2018dnnbuilder} on a ZC706 FPGA with 900 DSPs with results shown in Table \ref{tab:edd-net-3}. By evaluating in the same ImageNet dataset using the same 16-bit data quantization, 
EDD-Net-3 achieves higher throughput (1.45$\times$) and accuracy by taking advantage of the co-design strategy compared to the VGG accelerator generated by DNNBuilder.

\section{Conclusions and Future Directions}
\label{sec:conclusion}

In this work, we provided a comprehensive discussion on effective algorithm-accelerator co-design approaches.
Three leading works were highlighted together with experimental results demonstrating that co-design approaches are effective in delivering hardware-friendly DNN algorithms as well as high performance hardware accelerators.
We reemphasize the importance and efficacy of algorithm-accelerator co-design, and call for more research activities in this demanding area.
Despite the remarkable achievements, there are still many open problems that would require new solutions. 

\textbf{Co-design for distributed AI.}
While AI technologies and applications have been advancing rapidly, AI algorithms are evolving from a centralized manner (cloud AI) to a distributed manner (edge AI),
where the algorithms and distributed accelerators will work collaboratively.
Such a new paradigm can largely alleviate the massive data storage and computation burden.
The emerging privacy and security requirement of the AI solutions also brings in new design challenges.
To adapt to such a paradigm shift,
novel co-design methodologies are required for distributed AI development and deployment.

\textbf{Co-design for heterogeneous and large scale AI.}
Another trend of future AI is heterogeneous and large scale.
For example in smart city, the heterogeneity resides in all aspects including data, algorithms and devices,
such as traffic planning, crowd monitoring, public health care, security, economy, and urban planning.
In addition to heterogeneity, such applications may also involve millions or billions of edge nodes at an extremely large scale.
To this end, novel co-design methodologies are essential to address these challenges and push AI to a new phase of real-world applications.

\textbf{Co-design for emerging AI technologies.}
Both AI algorithms and devices are emerging dramatically.
For example, the rapid development of process-in-memory technologies and the neuromorphic computing technologies
have brought in immense opportunities.
Moreover, the recent achievements of quantum computing have also opened a new era of AI development.
Towards these interesting yet challenging new directions of future AI, revolutionary new design methodologies for software and hardware are mandatory.

\section*{Acknowledgment}

This work is supported in part by the IBM-Illinois Center for Cognitive Computing System Research (C3SR) -- a research collaboration as part of IBM AI Horizons Network,
and Campus for Research Excellence and Technological Enterprise (CREATE) programme in Singapore.

\bibliographystyle{unsrt}
\bibliography{ref}

\end{document}